\documentclass{article}
\usepackage{spconf,amsmath,bm,graphicx}
\usepackage{booktabs}
\usepackage{spbmark}
\usepackage[backend=biber,sorting=none, style=ieee]{biblatex}
\title{A META LEARNING SCHEME FOR FAST ACCENT DOMAIN EXPANSION IN Mandarin SPEECH RECOGNITION}
\name{Ziwei Zhu, Changhao Shan\sthanks{The author contributed equally to this work.}\sthanks{Work done in Tencent.}, Bihong Zhang\super{$\dagger$}, Jian Yu\super{$\dagger$}}
\address{Input Division, Tencent PCG\\
alicezzhu@tencent.com}
\begin{document}
\maketitle
\begin{abstract}

Spoken languages show significant variation across mandarin and accent. Despite the high performance of mandarin automatic speech recognition (ASR), accent ASR is still a challenge task. In this paper, we introduce meta-learning techniques for fast accent domain expansion in mandarin speech recognition, which expands the field of accents without deteriorating the performance of mandarin ASR. Meta-learning or learn-to-learn can learn general relation in multi domains not only for over-fitting a specific domain. So we select meta-learning in the domain expansion task. This more essential learning will cause improved performance on accent domain extension tasks. We combine the methods of meta learning and freeze of model parameters, which makes the recognition performance more stable in different cases and the training faster about 20\%. Our approach significantly outperforms other methods about 3\% relatively in the accent domain expansion task. Compared to the baseline model, it improves relatively 37\% under the condition that the mandarin test set remains unchanged. In addition, it also proved this method to be effective on a large amount of data with a relative performance improvement of 4\% on the accent test set.

\end{abstract}

\begin{keywords}
accent domain expansion, meta-learning, accent speech recognition, fast training 
\end{keywords}

\section{Introduction}
\label{sec:intro}

Accents are produced on the basis of a mandarin with the change of some facts, such as initial consonants, finals, tones and rhythm as well. Due to regional distinctions and different language backgrounds, accents often appear in daily conversations. Despite the high performance reported by state-of-the-art mandarin automatic speech recognition (ASR) systems, the accent ASR is not ideal. The common method of transferring from mandarin ASR to accent ASR is domain adaptation, which trains additional models to obtain the best accent domain performance. However, in industry, we pay more attention to one model with common domain by domain expansion, which is more simple and effective.

Domain expansion is first proposed in \cite{2019Domainexpansion} , which is based on DNN based ASR models. End-to-end (E2E) model \cite{ctc_aed,wenet} is the mainstream model and we hope to explore methods based on E2E model for accent domain expansion in mandarin ASR. There are many related researches on domain expansion. Fine-tuning, retraining the model using a dataset of all domains, is a simple and easy way to use, which directly learns the tasks of the mandarin and accent speech recognition. However, it needs a large scale dataset and trains slowly.\cite{2019Domainexpansion,DAT,DAT_amazon,2020Domain_KLD} focus on applying accent invariant acoustic features and adaptation methods to allow the model to accommodate accented speech without the performance degradation of mandarin ASR. However, the learned features and models only apply to the domains in the training set and the essence of features and models for all domains is not learned. Meta learning has been introduced to cross-accented speech recognition for fast adaptation to unseen accents in \cite{metalearningunseendomains}. We want to explore meta-learning methods for known accents in common ASR. 

In this paper, we introduce meta-learning techniques for accent domain expansion in mandarin ASR. Different from fine-tuning and accent invariant feature learning, the reference to meta-learning enables the model to learn methods for speech recognition in different scenarios without reducing the performance of common scenarios, which is a more essential learning. Inspired by the work of \cite{tomanek2021residual} that, similar adaptation gains is achieved by only updating a tiny fraction of the model parameters, we combine the methods of meta learning method and freeze of model parameters to make the model more stable and the training faster. In a conclusion, our major contribution includes:
\begin{itemize}
\item[1)] The work is used for an accent domain expansion task on common ASR instead of domain adaptation.
\item[2)] we introduce meta learning for the accent domain expansion task, which is more effective and essential. 
\item[3)] Fast training is realized by adding a freeze of model parameters to the meta learning.
\item[4)] We have proved the work on a large-scale data trained model. 
\end{itemize}
The rest of the paper is organized as follows. Section 2 summarizes related work and Section 3 describes the proposed meta-learning method for accent domain expand task. Experiments and analyses are provided in the Section 4. Section 5 concludes the paper.

\begin{figure}[t]
\centering
\includegraphics[scale=0.4]{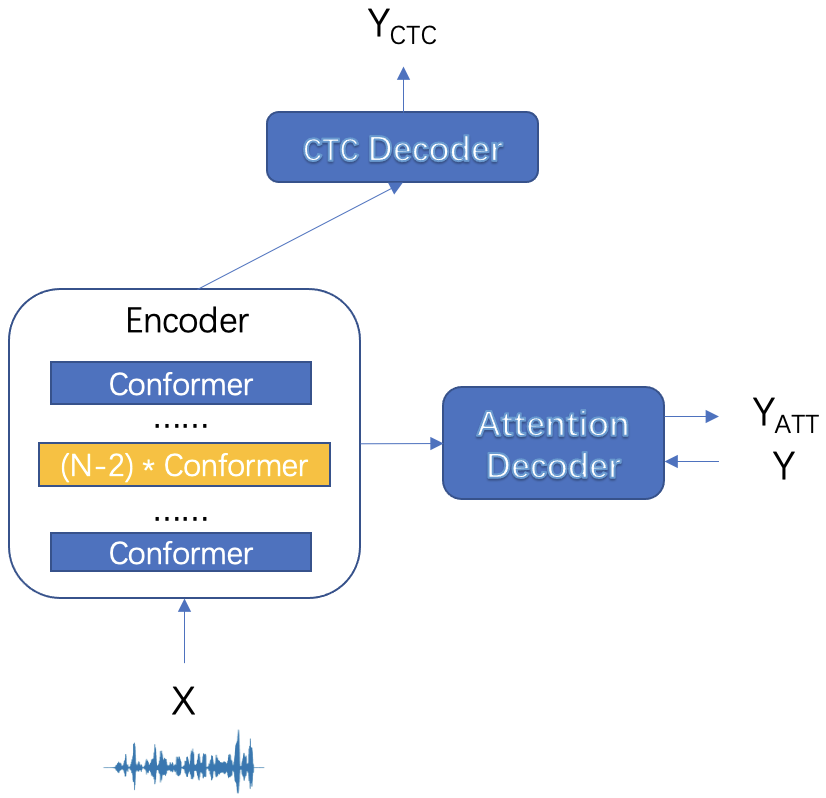}
\caption{A joint CTC/AED speech recognition system}

\end{figure}

\section{Related Work}
\subsection{Accent Domain Expansion}
Domain expansion is a type of transfer learning. Different from domain adaptation, the goal of domain expansion is to modify the model such that is performs well for original and new domains. The fundamental difficulty of domain expansion is to preserve the functionality of the original domain model in the process of new domain learning, which is minimal forgetting effect. Many approaches have been proposed to deal with the forgetting problem in domain expansion. In this paper, we mainly solve the problem of accent domain expansion on 
common ASR.

\subsection{Methods for Accent Domain Expansion and Adaptation}
To deal with the problem of accent domain expansion and adaptation, researchers mainly study the several techniques: 1) weight constraint adaptation (WCA) \cite{2019Domainexpansion}: keeping the model parameters close to the original model parameters by limiting weight; 2) KL-divergence (KLD) \cite{2020Domain_KLD, TS}: restricting the KL-divergence between the original and the adapted model output distributions; 3) meta learning \cite{metalearningunseendomains,vinyals2016matching,snell2017prototypical,rusu2018meta}: learn to learn the accent speech recognition; 4) freeze of model parameters (FMP) \cite{tomanek2021residual}: only adjusting the accent related features and parameters; 5) exporting extra models \cite{AISpeech-SJTU_AID,JMAAMSR_AID,Xinjiang_AID}: export accent identify model to help accent speech recognition; 6) the hybrid of these methods. For a common ASR, the combine of WCA and KLD performs best because it preserves the parameters of the original model as much as possible in the domain expansion process. In this paper, we explore learn to learn based methods for accent domain expansion tasks on E2E ASR.

\section{Methodology}
In this section, we first present the architecture of Conformer-based speech recognition, then the proposed method using meta-learning and freeze of model parameters for fast accent domain expansion on the common ASR is introduced.

\subsection{A Joint CTC/AED Model}
As Figure 1, we build our model using a joint CTC/Attention-based Encoder-Decoder(AED) model to learn to predict labels from the speech input. The joint CTC/AED model comprises three parts: 1) a shared chunk-based Conformer Encoder; 2) a CTC Decoder; 3) an Attention Decoder. 

The shared Encoder learns a hidden representation $H=(h_{u}|u=1,...,U)$ from a speech feature sequence $X=(x_{t}|t=1,...,T)$. Then, The CTC Decoder and Attention Decoder transform the encoder output $H$ to the CTC output and AED output respectively. Finally, the hybrid loss of CTC and AED is: 
\begin{equation}
    \label{eq1}
    L_{CTC-AED}(X,Y) = \lambda L_{CTC} + (1- \lambda)L_{AED}
\end{equation}
where $\lambda$ is a hyper-paramter: $0<\lambda<1$.

In the inference stage, we use beam search on WFST-based CTC decoding like \cite{sogou_WNARS}, in which a n-gram word-level LM is integrated by a WFST to generate n-best hypotheses during CTC decoding.

\subsection{Accent Domain Expansion Via Meta-Learning}
Meta-learning can learn general relation in multi domains not only for over-fitting a specific domain\cite{hsu2020meta,mi2019meta}. Based on the joint CTC/AED model, we apply model-agnostic meta-learning (MAML) \cite{pmlr-v70-finn17a-MAML} to effectively learn from the mandarin and accent speech and expand accent domain without deteriorating the performance of mandarin ASR. We denote our ASR model as $f_{\theta}$ parameterized by $\theta$, the common speech data as $G$, and the accent speech data as $A=\{A_1,A_2,...,A_n\}$. The total training data is $I=\{G,A\}$. We split $I$ into training set $I^{tra}$ and support set $I^{sup}$. It need to be noted that different samples may appear in different sets in different epoches.

For every iteration $i$ in one epoch, we select some sample batches from $I^{tra}$ as training set $I^{tra}_i$. The gradients are computed on the set $I^{tra}_i$ by the model $\theta$, and then the parameters $\theta$ can be updated as $\theta'$.

\begin{equation}
    \label{eq2}
    \theta' = \theta - \alpha \nabla_{\theta}I^{tra}_i(f_{\theta})
\end{equation}

where $\alpha$ is the learning rate for training set. After the learning of training set, we select samples from $I^{sup}$ as support set $I^{sup}_i$. The gradients $\nabla_{{\theta}^{'}}$$ I^{sup}_i $$(f_{{\theta}^{'}} )$ of support set is calculated by the model $\theta{'}_i$.

At the end of the iteration, the model $\theta$ is updated by $\nabla_{{\theta}^{'}}$$ I^{sup}_i $$(f_{{\theta}^{'}} )$, which learns to learn the mandarin and accent speech recognition method. 
\begin{equation}
    \label{eq3}
    \theta \leftarrow \theta - \beta \nabla_{{\theta}^{'}}I^{sup}_i(f_{{\theta}^{'}})
\end{equation}

The essential difference between fine-tuning and meta-learning is the way of gradients updating. If he gradient learned from one set can be used for other sets, we think it is meta-learning which learns the method for recognition. 

\subsection{Combination of Freeze of Model Parameters and Meta-Learning for Fast Accent Domain Expansion}
Meta learning is a steeper process, meaning greater uncertainty. To prevent excessive vibration of the model, we freeze the part of the model parameters that has little impact on the difference between accent and mandarin. The yellow blocks in Figure 1 represent the frozen model parameters, which are less relevant to acoustic characteristics and text labels and are considered not critical. Correspondingly, the blue parts refer to the updated parts in the system. Only training a fraction of parameters not only makes the sensitive cases in the general data change very little, but also it can improve the training speed. 

It is an interesting question that, if not limited by the model structure of speech recognition, how to select the updated and frozen parts? We believe that the layers close to the acoustic features and the layers close to the text label in the Encoder should be updated when training.

\section{Experiments} 
\subsection{Training Description}
\subsubsection{Datasets}
For the training set, we used a Chinese speech interaction data for the experiments in Table 1. We collected 10k hours speech data as Mandarin dataset and 3k hours accent speech as Accent dataset in various ways. These data have been manually labeled and verified. The Accent dataset covers about 11 provinces (including Guangdong, Fujian, Jiangsu, Chongqing, Shanghai, Hunan, Henan, Jiangxi, Shanxi, Zhejiang and Guizhou). All dataset consists of Accent dataset and Mandarin dataset, while Accent+ dataset comprises Accent dataset and 3.6k hours speech data selected from Mandarin dataset. In addition, we use 50,000-hour corpus to verify the proposed method on the large-scale data trained model.

\begin{table}[ht]
\caption{The summary of training datasets.}
\vspace{5pt}
\begin{center}
\begin{tabular}{cc}
\toprule 
Dataset & Data Size  \\
\midrule 
Accent & 3.6kh accent speech\\
Mandarin  & 10kh mandarin speech\\
All & 3.6kh accent + 10kh mandarin\\
Accent+ & 3.6kh accent + 3.6kh mandarin\\
50,000-hour & 50kh mandarin\&accent speech \\
\bottomrule 
\end{tabular}
\end{center}
\end{table}

\begin{table}[ht]
\vspace{-20pt}
\caption{The task of accent domain expansion.}
\vspace{5pt}
\begin{center}
\begin{tabular}{cccc}
\toprule 
Name & Data & Mandarin & Accent \\
\midrule 
Baseline-1 & Mandarin & 8.72 & 13.96\\
Baseline-2 & All & 8.54 & 9.03 \\
\quad +Fine-tune & Accent & 9.54 & 8.43 \\
\bottomrule 
\end{tabular}
\end{center}
\vspace{-10pt}
\end{table}

For the testing set, the average of results of 11 provinces accent testing sets computes the Accent testing set and the Mandarin testing set is also an average of different testing sets. Accent-2 and Mandarin-2 are testing sets for large scale ASR system, which are the average of testing sets with different scenarios.

\subsubsection{Model setup}

We used the CTC/AED based Conformer model in all setups, in which the Encoder consists of 12 causal Conformer layers and Attention Decoder consists of 4 transformer layers. The feed-forward dimension was 2048 and the attention dimension was 256. The Encoder used a multi-head of 6 while the Decoder used a multi-head of 4. 

The input acoustic features used for all experiments were 71-dimensional log-mel filter-banks computed on 25ms window with a 10ms shift. Down sampling was performed through two convolution layers with a factor of 4. Acoustic modeling units were the Chinese characters and English word pieces, which were totally 16776. Adam optimizer and Noam \cite{Attention} learning rate schedule are used with 25,000 warm-up steps at the process of training. We also employed label smoothing and dropout regularization of 0.1 to prevent over-fitting. Moreover, the weight of the CTC loss was set to 0.3. We trained all models with 8 NVIDIA TESLA V100 GPUs, using the ESPnet \cite{ESPNET} toolkit with the PyTorch backend.
We report character error rate (CER) for all experiments.

\subsection{Results and Discussions}

\subsubsection{The baseline of accent domain expansion }

Baseline-1 in Table 2 is the initial model of our experiment, which is trained with a Mandarin dataset. Baseline-2, which uses mandarin data and accent data, not only improves the performance of Accent testing set relatively about 35\%, but also the effect of the Mandarin test set increases 2\% compared to Baseline-1. This shows that accent in the Accent dataset is not so strong that the adding of accent data can also improve the effect of the Mandarin testing set. If we only use the Accent dataset to fine-tune Baseline-2, performing the Mandarin test set will decline by 12\% while the accent performance was improved 6\%. We hope to exploit only the data of the new accent data to build a stronger model for all domains, so we explore some methods and introduce the proposed method as follows.

\begin{table}
\caption{The performance of meta-learning and other domain expansion methods.}
\begin{center}
\begin{tabular}{cccc}
\toprule 
Model & Data & Mandarin & Accent \\
\midrule 
FT & Accent & 9.54 & 8.43 \\
KLD & Accent & 8.60 & 8.71\\
KLD & Accent+ & 8.58 & 8.96 \\
WCA & Accent & 8.47 & 9.66 \\
FMP & Accent & 8.52 & 8.69 \\
MAML & Accent & 8.82 & \pmb{8.61} \\
MAML & Accent+ & \pmb{8.46} & 8.69\\
\bottomrule 
\end{tabular}
\end{center}
\vspace{-20pt}
\end{table}

\subsubsection{Comparison of meta-learning and other domain expansion methods}

Table 3 is a comparison of different methods of accent domain expansion. Although only the accent data is used, performing these four methods on the Mandarin testing set is better than fine-tuning, while the improvement of accent performance is not as good. 

The WCA method has the best performance on the Mandarin test set but the worst performance on the Accent testing set. The FMP and KLD methods both have good performance on the Mandarin testing set and the Accent testing set. FMP helps to update the accent variant parameters in the model when learning accent invariant features. KLD uses the mandarin model as the teacher model, which gives consideration to performing mandarin when learning accent. Because of the variance between accent and mandarin, FMP is slightly better than KLD on Mandarin Accent test sets and has a relative improvement of 3.4\% compared to Baseline-2. 

\begin{table}
\begin{center}
\caption{The performance of the combination of meta-learning and other methods for accent domain expansion.}
\begin{tabular}{cccc}
\toprule 
Model & Data & Mandarin & Accent \\
\midrule 
MAML & Accent+ & 8.46 & 8.69\\
MAML + FMP & Accent & 8.48 & 8.77 \\
\bottomrule 
\end{tabular}
\end{center}
\vspace{-10pt}
\end{table}

The above methods preserve the functionality of the original domain model in the new domain learning process. However, MAML needs to learn accent speech recognition method and common speech recognition method simultaneously. Therefore, we have added the same proportion of mandarin data based on accent data. The MAML method with Accent+ outperforms the MAML with Accent with a relative improvement of 4\% on Mandarin test set and a slight decline on Accent test set. For methods that are not inferior to Baseline-2 in Mandarin test set, MAML with Accent+ training is the best choice for Mandarin and Accent test sets, which increases the effect 3.7\% relatively.

\subsubsection{Fast training for accent domain expansion}

To make the model training more stable and faster, we combine the MAML and FMP in Table 4. The combination of MAML and FMP performs worst on Accent test set and best on Mandarin test set compared to single models with a speed improvement of 60\%, in which 20\% is because of the freeze of parameters and 50\% is because of the less training data. Meanwhile, the MAML+FMP model outperforms 3\% relatively than other methods in the accent domain expansion task and has a relative improvement of 37\% than Baseline-1 under the condition that keeps the CER of Mandarin test set unchanged.

\begin{table}
\caption{The performance of meta-learning for accent domain expansion on large amount data.}
\begin{center}
\begin{tabular}{ccc}
\toprule 
Model &  Mandarin-2 & Accent-2 \\
\midrule 
Baseline & 7.71 &  7.5 \\
MAML+FMP & \pmb{7.52}  & \pmb{7.2} \\
\bottomrule 
\end{tabular}
\end{center}
\vspace{-20pt}
\end{table}

\subsubsection{Comparison of meta-learning and other domain expansion methods}
We have verified the optimal configuration on a large amount of data in Table 5. It shows that the combination of MAML and FMP methods for accent domain expansion also applies to the large-scale speech recognition system, which achieves a relative improvement of 4\% on Accent-2 test set and 2\% on Mandarin-2 test set.

\section{Conclusion}

In this paper, we introduce meta learning techniques for accent domain expansion in common ASR. Meta-learning can learn general relation in multi domains not only for learning a specific domain. So meta-learning is selected in the domain expansion task. FMP is added to the MAML training, which makes the model more stable and the training faster by 60\%. The proposed approach significantly outperforms 3\% relatively than other domain expansion methods and has a relative improvement of 37\% than baseline, while the effect of the mandarin test set remains unchanged. In the future, we will investigate an improved version of the proposed domain expansion method for a wider range of domains, such as denoising and so on.

\printbibliography
\end{document}